\documentclass[12pt,a4paper,english,amssymb,nofootinbib,superscriptaddress]{revtex4-2}
\usepackage{lmodern}
\usepackage{lmodern}
\usepackage[T1]{fontenc}
\usepackage[latin9]{inputenc}
\usepackage{color}
\usepackage{babel}
\usepackage{verbatim}
\usepackage{amsmath}
\usepackage{amssymb}
\usepackage{esint}
\usepackage{graphicx}
\usepackage{subfigure}
\usepackage{tikz}
\usetikzlibrary{cd}

\usepackage[unicode=true,pdfusetitle,
 bookmarks=true,bookmarksnumbered=false,bookmarksopen=false,
 breaklinks=false,pdfborder={0 0 1},backref=false,colorlinks=true]
 {hyperref}
\hypersetup{
 linkcolor=blue,citecolor=blue}

\makeatletter

\pdfpageheight\paperheight
\pdfpagewidth\paperwidth

\@ifundefined{textcolor}{}
{%
 \definecolor{BLACK}{gray}{0}
 \definecolor{WHITE}{gray}{1}
 \definecolor{RED}{rgb}{1,0,0}
 \definecolor{GREEN}{rgb}{0,1,0}
 \definecolor{BLUE}{rgb}{0,0,1}
 \definecolor{CYAN}{cmyk}{1,0,0,0}
 \definecolor{MAGENTA}{cmyk}{0,1,0,0}
 \definecolor{YELLOW}{cmyk}{0,0,1,0}
}

\usepackage{babel}
\usepackage{babel}
\usepackage{babel}
\usepackage{babel}
\usepackage{babel}
\usepackage{babel}
\@ifundefined{definecolor}{color}{}
\usepackage{babel}
 \newcommand{\Hquad}{\hspace{0.3em}}
 \newcommand{\HHquad}{\hspace{0.1em}}

\usepackage{latexsym}\usepackage{bm}

\makeatother

\begin{document}
\title{A holographic description on half planes and wedges for $\mathcal{N}=1$ SUSY $BF$ theory in $2D$
}

\author{Cagdas Ulus Agca}
\email{ulus.agca@metu.edu.tr}

\affiliation{{\small{}Department of Physics,}\\
 {\small{}Middle East Technical University, 06800 Ankara, Turkey}}

\begin{abstract}
\noindent $BF$ theory is a topological field theory that appears in different parts of theoretical physics and one of its important uses is in lower dimensional holography setting. While it can be defined as a dimensional reduction of $3D$ $CS$ theory, it is also equivalent to $JT$ gravity. Moreover, further holographic settings relate $BF$ theory to a particle on group theory. Here, we reconsider this "simplest holography" construction as $SL(2,\mathbb{R})$ invariant particle on group theory and extend the web of dualities diagram in terms of a holographic description of the $2D$ $\mathcal{N}=1$ $BF$ theory on half-plane and find its $1D$ particle on group description as $\mathcal{N}=1B$ supermultiplet superconformal quantum mechanics. Moreover, we provide wedge space holography-type construction to achieve codimension $2$ holography and show the web of dualities diagram also closes diagonally for $BF$ and $CS$ theories. For $BF$ theory it leads to a complex $1D$ particle on group mechanics on the face of the wedge, a $0D$ quantum mechanics on the boundary theory of the faces, and its supersymmetric extension is a supermultiplet containing complex boson and complex scalar of the Landau-Ginzburg type. Upon this construction boundary theory also realizes a global $U_j(1)$ invariance which extends the total symmetry group. To have a more complete picture we also included wedge space holography for $3D$ CS theory as an appendix and showed its codimension 2 holography corresponds to a $1D$ particle on group we found earlier.
\end{abstract}
\maketitle

\section{Introduction}

After Maldacena realized a duality between supergravity theory in $5$ dimensional anti-de Sitter space and its boundary theory $4$ dimensional $\mathcal{N}=4$ super Yang-Mills theory \cite{Maldacena:1997re}, quantum gravity research was lit through the lens of the holographic principle \cite{tHooft:1993dmi}. Building up models of quantum gravity where gravity has non-zero bulk degrees of freedom has had its problems since showing the correspondence from both sides was sometimes impracticable due to high oscillations while integrating inside the bulk degrees of freedom by using the holographic dictionaries \cite{Gubser:1998bc,
Witten:1998qj}. Meanwhile, Witten \cite{Witten:1988hc,
Witten:2007kt} had judiciously chosen the gauge fields in terms of vielbeins of $3D$ gravity, which has no graviton in the bulk, showed a classical equality between non-abelian Chern-Simons theory and $3D$ gravity with cosmological constant and then showed the equality holds at the quantum level. Followed by the seminal paper of Brown-Hanneoux, non-abelian CS theory equivalent to $Sl(2,\mathbb{R})$ $WZW$ model on the boundary. This theory can also be seen as a conformal field theory through its asymptotical symmetry structure \cite{Brown:1986nw,Donnay:2016iyk}, abbreviated as $AdS_3/CFT_2$. The solvable nature of $CFT_2$ and its symmetry structure infinitely constrains the partition functions of quantum gravity. Hence, the search for lower dimensional holography through $D<3$ $CFT$s and topological field theories has begun. Through the lower dimensional considerations of gravity such as $JT$ gravity \cite{Jackiw:1984je,
Teitelboim:1983ux}, and its flat space limit \cite{Afshar:2019axx,Afshar:2019tvp}, Dilaton Gravity \cite{Grumiller:2002nm}  and $CGHS$ gravity \cite{Grumiller:2006xz}. One can see \cite{Iliesiu:2019xuh} has shown that the similar change of variables in $3D$ gravity and $3D$ non-abelian Chern-Simons theory works for lower dimensional gravity theories. Through this lens, $BF$-type theories found their place in lower dimensional holography research on top of topological superconductors \cite{Balachandran:1992qg}. On manifolds with boundaries, the $BF$ theory realized as gapped non-chiral edge states \cite{Hansson:2004wca}. Moreover, $BF$-like theories are shown as toy spin-foam models \cite{Baez:2006sa}. Recently, it has been shown that coset $CGHS$ supergravity's gauge theoretic formulation encapsulates $BF$ theory
\cite{Afshar:2022mkf}. Furthermore, $BF$ theories, while equivalent to $JT$ gravity, have been shown they also obey the $1D$ particle on group \cite{Grumiller:2020elf}
quantum mechanics prescription \cite{Janik:2018dll} and their $T\Bar{T}$ deformations \cite{Ebert:2022ehb} when external Schwarzian boundary terms are added to the action. These links between low-dimensional gravity theories and their corresponding gauge theories led to the building up a web of dualities in terms of low-dimensional holography. It is crucial to understand the realm of holographies and their possible supersymmetric extensions.
In this paper, we will consider  $\mathcal{N}=1$ supersymmetric extension of $2D$ $BF$ theory \cite{Constantinidis:2012jk}, which is manifested to be invariant under $\mathcal{N}=1$, by reformulating in the superspace formalism of $2D$ topological $BF$ theory with matter coupling \cite{Leitgeb:1999ep}. We will extend the web of dualities diagram considering $2D$ $BF$ theory as a supersymmetric theory and show its topological holography on the boundary as a $1D$ particle on group superconformal quantum mechanics. Furthermore, we consider $2D$ $BF$ on wedge space which realizes codimension 2 holography in the sense of \cite{Ogawa:2022fhy} where we find complex $1D$ particle on group and theory while its supersymmetric extension realized a Landau-Ginzburg type supermultiplet. Finally to have a more complete picture of the web of dualities diagram we provide a wedge space holography construction for $3D$ $CS$ theory and show that 
diagonal arrows can also be put into the diagram.

The structure of this paper is as follows: In section (\ref{section2}), we give a brief review of $BF$ theories in generic dimensions. In (\ref{nonsusyholography}), we will review the holographic correspondence of non-abelian $BF$ theory with proper boundary conditions on half-plane and for coherence with conformal quantum mechanics we choose particle on $Sl(2,\mathbb{R})$ group. Furthermore, in (\ref{susyformalism}), we define $\mathcal{N}=1$ superspace formalism in the sense of \cite{Constantinidis:2012jk,Leitgeb:1999ep}. In the last section (\ref{section4}), we find its particle on group quantum mechanics correspondence on the boundary and show that it is invariant under superconformal transformations with its supergauge group and finally in (\ref{wedgesection}) we give its wedge space holography construction.

\section{A brief review of $BF$ Theory}\label{section2}

As one of the simplest topological field theory models, the $BF$ theory found its usage in various areas from lower dimensional holography, string theory, condensed matter physics, spin-foam models and loop quantum gravity, conformal field theories, etc. This section will review this influential theory in $2D$ for readers unfamiliar with its basic construction. The author does not claim the novelty of the basic construction presented below.

In general dimensions, $BF$ theory is defined by choosing a Lie group $G$ acting on a $D$-dimensional oriented manifold M \cite{Baez:1999sr}. The corresponding Lie algebra $\mathfrak{g}$ has a bi-linear form $\langle.,.\rangle$, which we will take as a Killing form by choosing a semi-simple Lie group on M. To write a meaningful action, let there be a connection $A$ on the principal bundle of $G$ and a $(D-2)$-form $B$ transforming under the adjoint representation of the principal bundle living on our oriented manifold $M$. Hence, while the connection is $\mathfrak{g}$-valued 1-form, its curvature $F$ is a $\mathfrak{g}$-valued 2-form. Then, the action principle can be written as:
\begin{equation}
    S= \int_\text{M} Tr( \Hquad B \wedge F \Hquad).
\end{equation}
Variation of this action leads to 
\begin{flalign}
    \delta_\text{A} S=&\int_\text{M} Tr( \Hquad \delta B \wedge F \Hquad+ \Hquad  B \wedge \delta F \Hquad)\notag,\\
    = &\int_\text{M} Tr( \Hquad \delta B \wedge F \Hquad+ \Hquad  (-1)^{n-1}d_\text{A} B \wedge \delta A \Hquad)=0,
\end{flalign}
which gives
\begin{flalign}
    F=&0,\quad\quad\quad d_\text{A}B=0,
\end{flalign}
where $d_\text{A}$ is the exterior gauge covariant derivative. Here, we write $F$ as the exterior covariant derivative of $A$, interchanged the variation and $d_\text{A}$ following an integration by parts to write down field equations. One can immediately see that solutions $F$ lead to a set of pure gauge transformations that purely depend on elements of semi-simple algebra. It hints at the fact that there are no local degrees of freedom or a metric dependence in that sense; hence, the $BF$ theory is a topological field theory.\footnote{$BF$ model's deep connections with lower dimensional holography can be understood from this phenomena since $AdS_3$ gravity also has no local degrees of freedom.}

\section{Holography of BF theory}\label{nonsusyholography}
In this section, we will try to motivate how to relate quantum mechanical models on group manifolds with topological field theories under holographic construction. In what follows\footnote{We thought it best suited to review the "simplest holography" approach by Janik \cite{Janik:2018dll}, which we believe includes easy-to-follow calculations. Afterwards, we will present its $\mathcal{N}=$1 SUSY extension.}, from now on we will take a $2$-dimensional half-plane with a boundary $\text{M}=\{(t,x): x\geq0\}$ and $G$-valued connections on $2D$ $BF$ model. One can realize that in $2$ dimensions, the $B$-field is a Lie algebra valued $0$-form, i.e. $\mathfrak{g}$-valued scalar field $\phi^i$. Hence, one can rewrite the $LSZ$ action without matter coupling as \cite{Leitgeb:1999ep}:
\begin{flalign}
S_{BF}=&\int_\text{M} d^2 x \Hquad\frac{1}{2}\epsilon^{\mu\nu}\phi^iF^i_{\mu\nu},\\
=&\int_\text{M} d^2 x \Hquad \phi^i\Hquad(\partial_tA^i_x-\partial_x A^i_t+f^i_{jk} A_t^jA_x^k).
\end{flalign}
Since we have a boundary at $\{x=0\}$, we should support our action with a Schwarzian-type boundary condition \cite{Ebert:2022ehb}. It will satisfy the action principle. 
Let us first take the variation
\begin{flalign}
\delta_\text{A} S_{BF} =&\int_\text{M} d^2 x \Hquad \phi^i\Hquad(\partial_t\HHquad\delta_\text{A}A^i_x-\partial_x \HHquad\delta_\text{A}A^i_t+f^i_{jk}\HHquad \delta_\text{A}A_t^jA_x^k+f^i_{jk}\HHquad A_t^j\HHquad\delta_\text{A}A_x^k).
\label{variedBFtheory}
\end{flalign}
One can take integration by parts for both $t$ and $x$ derivatives and can assume that the total derivative in the $t$-direction gives a regular boundary at infinity variation after taking the integral. Hence, it has no contribution to the boundary. However, in $x$-direction, we have non-trivial on-shell boundary action 
\begin{equation}
   = \text{eom}+\delta_\text{A}S_{\text{boundary}},
\end{equation}
where "eom" is short for the equation of motion, the boundary action can be written as:
\begin{equation}
    \delta_\text{A}S^{\text{on-shell}}_{\text{boundary}}=\int_\text{x=0} dt \Hquad(\phi^i\delta_\text{A}A^i_t).
\end{equation}
Now, concerning this boundary, one can redefine the action itself with an external boundary cancellation term of the Schwarzian type; upon variation, it manages to cancel the non-zero on-shell boundary contribution. 
\begin{equation}
     \delta_\text{A}S^{\text{on-shell}}_{\text{boundary}}=\int_\text{x=0} dt \Hquad(\phi^i\delta A^i_t-\phi^i\delta\phi^i).
     \label{susyanalogueofschwarzianboundarycorrection}
\end{equation}
For this to be zero, we have to equate $A^i_t$ and $\phi^i$ on the boundary.
\begin{equation}
    A^i_t\vert_{\text{x}=0}=\phi^i\vert_{\text{x}=0}.
    \label{nonsusyboundarycondition}
\end{equation}
Hence, our boundary action becomes 
\begin{flalign}
    \delta_\text{A}S_{\text{boundary}}=&\int_\text{x=0} dt \Hquad(A_t^i\delta A^i_t)
        =\int_\text{x=0} dt \Hquad\delta(\frac{1}{2}A_t^i A^i_t).
        \label{totalvariatonofAt}
\end{flalign}
Now, we rely on the topological nature of the $2D$ $BF$ theory. Since $F=0$, the transformation $A'=gAg^{-1}+gdg^{-1}$ can be set as $A=0$. Hence, the pure gauge solutions reduce to $A'=-g^{-1}\partial_\mu g$, and primed indices can be redefined as unprimed. Now let us substitute the pure gauge solutions into \eqref{totalvariatonofAt}
\begin{flalign}
   S_{\text{boundary}}=&\int_\text{x=0} dt \Hquad\frac{1}{2}(-g^{-1}\partial_t g) (-g^{-1}\partial_t g)),
        \label{totalvariatonofAt1}
\end{flalign}
 the integrand can be recast as
\begin{equation}
    (-g^{-1}\partial_t g) (-g^{-1}\partial_t g)=g_{ab}(\alpha)\dot{\alpha}^a\dot{\alpha}^b,
\end{equation}
 on the boundary we can $\lim_{x\to0} \dot{\alpha}^a=\dot{x}^a$
and if we substitute into \eqref{totalvariatonofAt1}
\begin{flalign}
    S_{\text{boundary}}=&\int_\text{x=0} dt \Hquad\frac{1}{2}g_{ab}(x)\dot{x}^a\dot{x}^b. 
    \label{janikwithgeneralgroup}
\end{flalign}
 We can choose our group manifold that our fields are defined on as flat; hence, the $\mathfrak{g}$-dependent metric on the group manifold can be trivialized to reproduce the results in Janik's paper. Thus, the action corresponds to a $1D$ particle on group quantum mechanics in general. We have a well-known holography between $2D$ $BF$ theory/$1D$ free particle quantum mechanics \cite{Janik:2018dll}. From our perspective, it is better to choose the group manifold as $SL(2,\mathbb{R})$ since, in the end, we will work with its simplest supersymmetric extension. We believe that with that choice, we have a more complete and coherent picture on both $\mathcal{N}=0$ and $\mathcal{N}=1$ where on the boundary, their conformal and superconformal extensions can be seen more easily. Let us interpret this action as conformally invariant quantum mechanics by choosing our gauge group as $SL(2,\mathbb{R})$ \cite{Michelson:1999zf,
Britto-Pacumio:1999dnb}, which can take the role of a toy model for our supersymmetric case. Firstly, let us write Hamiltonian for the action given in \eqref{janikwithgeneralgroup}.

\begin{equation}
    H=\frac{1}{2} P^\dagger_a g^{ab}P_b,
\end{equation}
such that $P_a=g_{ab}\dot{q}^b$ is the corresponding conjugate momentum where $a,b$ run over the number of particles $\{1,2,\cdots,N\}$, according to \cite{Michelson:1999zf}, the conformal and superconformal nature of these theories manifest with homothety condition. Dilatations can be generated as

\begin{flalign}
    \delta_D q^a=\epsilon D^a(q),\quad\quad\quad\quad\quad
     \delta_D\Hquad t= 2\epsilon t,
\end{flalign}
with the generator $D=\frac{1}{2}D^aP_a+ \text{h.c}$ obeys
\begin{equation}
    [D,H]=-\frac{i}{2}P_a(\mathcal{L}_D g^{ab})P_b-\frac{i}{4}\nabla^2\nabla_a D^a.
\end{equation}
    Hence, dilatational symmetry exists if and only if 
\begin{equation}
    \mathcal{L}_D g^{ab}=2g_{ab},
\end{equation}
where dilatational vector field manifests homothety. One can generate rest of the $SL(2,\mathbb{R})$ algebra as
\begin{flalign}
    [D,K]=-2iK,\quad\quad\quad\quad [H,D]=-iD,
\end{flalign}
and $\mathcal{L}_D K=2K$. This gives a closed homothety in terms of $D_a\HHquad dq^a=dK\implies d(D_a\Hquad dq^a)=0$ which means one form $D$ is exact with closed homothety. Solving the closed homothety equation gives the form of the special conformal transformations $K=\frac{1}{2}g_{ab}D^aD^b$.\footnote{For single particle case, \cite{Britto-Pacumio:1999dnb} shows that from Hamiltonian perspective the spectrum is continuous however, in Virasoro form of $SL(2,\mathbb{R})$ one can produce quadratic Casimir operators and find the ground state with infinite tower of eigenvalues increasing with integer valued eigenvalues. } Through the choice of $SL(2,\mathbb{R})$ group, it shows that the $2D$ BF theory corresponds to $1D$ quantum mechanics on a group which can be expanded to admit a conformal structure, i.e conformal quantum mechanics. In the next section, we will generalize this perspective to SUSY BF theory and show its boundary theory is equivalent to superconformal quantum mechanics given in \cite{Michelson:1999zf} with the same choices they make on homothety of the algebra elements.

\section{$\mathcal{N}=1$ Superspace formalism of $2D$ $BF$ theory}\label{susyformalism}
We aim to find a similar correspondence while preserving the supersymmetry. To establish our notation, we will review $\mathcal{N}$=1 superspace formalism for $BF$ theory. Firstly, we should start by defining the Leitgeb-Schweda-Zerrouki ($LSZ$) action, which can be seen as a matter-coupled extension of the $2D$ $BF$ action \cite{Leitgeb:1999ep}. Authors \cite{Constantinidis:2012jk} realized that this model fits into superspace formalism with a set of rigid supersymmetry transformations such that the action is left invariant. We will take this point of view from now on and try to construct everything that is done for "simplest holography" and find its extension of $\mathcal{N}=1$  SUSY. Let us rewrite the action and add the matter terms.
\begin{equation}
S_{BF}=\int_\text{M} d^2 x \Hquad\frac{1}{2}\epsilon^{\mu\nu}\phi^iF^i_{\mu\nu}+\int_\text{M} d^2 x \Hquad \epsilon^{\mu\nu}(D_\mu B_\nu)^i\psi^i,
\end{equation}
where $\phi^i$ is a scalar field that acts as a Lagrange multiplier, $F^i_{\mu\nu}$ is $G$-valued curvature $2$-form of $A^i_\mu$ curvature $1$-form. $B^i_\mu$ is the vector field and $D_\mu$ is the covariant derivative of the connection $A^i_\mu$.

The $\mathcal{N}=1$ superspace realization of this action is given by \cite{Constantinidis:2012jk}. We will simply review the basic construction. For $\mathcal{N}=1$ supersymmetry to be acquired by the action, it needs to be generated by a nilpotent operator $Q$. If we have $\mathcal{N}=1$ supersymmetry, the superfields can be expanded concerning their odd or even Grassmann nature. Moreover, we set $\theta^2=0$, which makes our superspace $1$ dimensional with a unique coordinate $\theta$. For a superfield given in 
\begin{equation}
    \phi(x^\mu,\theta)=\phi_0+\theta\phi_1,
\end{equation}
where $x^\mu=\{t,x\}$ will be the spacetime coordinates and $\theta$ will be the odd Grassmann variable. Since $Q=\partial_\theta$ as the generator of transformations, 
\begin{flalign}
    Q\phi_0=\phi_1,\quad\quad\quad\quad Q\phi_1=0.
\end{flalign}
Now, we will define our superfields to elevate the LSZ action and left invariant under $\mathcal{N}=1$ supersymmetry transformations.
\begin{flalign}
   \Phi=\psi+\theta\phi,\quad\quad\quad\quad \mathcal{A}=A+\theta B.
   \label{superfieldsevenoddstructure}
\end{flalign}
 $QA=B\quad QB=0$ and $Q\psi=\phi\quad Q\phi=0$, leaves $LSZ$ action invariant. Moreover, $\psi$ is odd, $\phi$ is even Grassmann, which leads to making $\Phi$ as odd Grassmann. Also, $\mathcal{A}$ is even since the gauge field $A$ is even, and $B$ has to be odd to be multiplied with $\theta$ and become even. This choice of superfields constitutes the superspace action of $LSZ$.
\begin{equation}
    S_{\textbf{SBF}}= \text{Tr}\int_{\mathcal{M}} d\theta \Phi \mathfrak{F}[\mathcal{A}],
\end{equation}
supercurvature can be decomposed into $\mathfrak{F}=F+ \theta\mathbb{F}$ where $F$ is the non-abelian curvature $2$-form, and $\mathbb{F}$ is its superpartner.
\begin{flalign}
    F=dA+\frac{1}{2}[A,A]=0\quad\quad\quad\quad\quad\mathbb{F}=dB+[A,B]
\end{flalign}
where the new action is on the supermanifold with supercurvature of superconnection $1$-form. If we assume vanishing boundary conditions, the variation of this action leads to
\begin{flalign}
   \delta S_{\textbf{SBF}}=& \text{Tr}\int_{\mathcal{M}} d\theta\Hquad \left(\delta\Phi\HHquad \mathfrak{F}[\mathcal{A}]+\Phi\delta\HHquad \mathfrak{F}[\mathcal{A}]\right)
   =\text{Tr}\int_{\mathcal{M}} d\theta\Hquad \left(\delta\Phi\HHquad \mathfrak{F}[\mathcal{A}]+\delta\mathcal{A}\HHquad \mathcal{D}\Phi\right).
   \label{variationofSBF}
\end{flalign}
 Here, $\mathcal{D}= d+ [\mathcal{A},]$ is the supercovariant derivative, and the trace is over the Killing form of the semi-simple algebra. The stabilized action gives the equations of motion as:
\begin{flalign}
    F=&0,\quad\quad\quad D\phi-[B,\psi]=0,\notag\\
    \mathbb{F}=&0,\quad\quad\quad D\psi=0,
    \label{Superequationsofmotion}
\end{flalign}
 Since both curvature $2$ form and its superpartner vanish in the level of the equation of motion. The supercurvature is also flat.
\begin{equation}
    \mathfrak{F}=0.
\end{equation}
 The symmetry transformations of superfields are controlled by supergauge group $\mathbf{G}$ with associated lie algebra $\mathbf{g}$ with elements
\begin{equation}
    \mathcal{G}(x)=\mathcal{G}(\alpha,\beta)=e^{\alpha(x)+\theta\beta(x)},
\end{equation}
where $\alpha$ is Grassmann even and $\beta$ is Grassmann odd. Expanding in $\theta$ gives an exact result since $\theta^2=0$.
\begin{equation}
    \mathcal{G}(\alpha,\beta)=\alpha(x)+\theta\beta\vartriangleright g(\alpha),
\end{equation}
where $\beta\vartriangleright g(\alpha)=Qg(\alpha)$. The superconnection transforms under $\mathcal{G}\in\mathbf{g}$ as
\begin{equation}
    \mathcal{A}(x)\rightarrow \mathcal{A'}(x)=\mathcal{G}(x)\HHquad d \HHquad\mathcal{G}^{-1}(x)+\mathcal{G}\mathcal{A}(x)\mathcal{G}^{-1},
\end{equation}
and flatness condition on the supercurvature leads
\begin{equation}
    \mathcal{A}_\mu=-\mathcal{G}^{-1}\partial_\mu\mathcal{G}.
\end{equation}
\section{From $\mathcal{N}=1$ Super-$BF$ theory to $\mathcal{N}=1$ Super-Quantum Mechanics }\label{section4}
In this section, we will apply the ideas in \ref{nonsusyholography} with the invariance of supersymmetry on our action in \ref{susyformalism}. In the previous case, we didn't have a boundary. Now, on a manifold with boundary $\text{M}=\{(\text{t,x}):x\geq0\}$, we will have non-trivial boundary terms at $\{\text{x}=0\}$. 
\begin{equation}
    S_{\text{SBF}}=\int dt\int_0^\infty dx\int d\theta \Hquad \Phi_i\left(\partial_t \mathcal{A}^i_x-\partial_x \mathcal{A}^i_t+f^k_{ij}\mathcal{A}^i_t\mathcal{A}^j_x\right) ,
    \label{SUSYBFTheoryaction1}
\end{equation}
where its variation,
\begin{equation}
    \delta_{\mathcal{A}} S_{\text{SBF}}=\int dt\int_0^\infty dx\int d\theta\Hquad \Phi_i\left(\partial_t \delta\mathcal{A}^i_x-\partial_x \delta\mathcal{A}^i_t+f^k_{ij}\delta\mathcal{A}^i_t\mathcal{A}^j_x++f^k_{ij}\mathcal{A}^i_t\delta\mathcal{A}^j_x\right) ,
\end{equation}
integration by parts leads
\begin{flalign}
\delta_{\mathcal{A}} S_{\text{SBF}}=&\int dt\int_0^\infty dx d\theta\Hquad \Phi_i\left(\partial_t\left(\Phi\delta\mathcal{A}^i_x\right)-\partial_t\Phi_i\delta\mathcal{A}^i_t-\partial_x(\Phi_i\delta A^i_t)+\partial_x\Phi\delta\mathcal{A}^i_t\right.\\&\left.+f^k_{ij}\HHquad\delta\left(\mathcal{A}^i_t\mathcal{A}^j_x\right)  \right).\notag
\label{variationofSBF}
\end{flalign}
Let us choose $\Phi(\pm\infty,x,\theta)=\mathcal{A}(\pm\infty,x,\theta)=0$ hence total derivative $t$ integral vanishes in \eqref{variationofSBF}. Other terms under variation will behave as the equation of motion when the action is on-shell, and only the total $x$ derivative term will contribute to the boundary where $\{\text{x}=0\}$ based on the choice of how the fields behave at the boundaries. Upon integration of the total $x$ derivative, we can also assume the fields are regular at $\Phi(t,\infty,\theta)=\mathcal{A}(t,\infty,\theta)=0$ on the spatial infinity. In the end, we will have the action variation as:
\begin{equation}
    S^{\text{on-shell}}_{\text{Boundary-SBF}}=\int_{\text{x=0}} dt \int d\theta\Hquad\Phi_i(t,0,\theta)\delta\mathcal{A}^i_t(t,0,\theta).
\end{equation}
The total derivative term on $x$ had a minus sign; however, $x$ integral is from 0 to $\infty$; hence, the minus sign becomes plus. This is the non-trivial remnant of the action for a manifold with a boundary. To satisfy the action principle, one should return the original action and add a correction term of the Super-Schwarzian type. However, at the level of superfields, the boundary correction term of the type $\int_{\text{x=0}} dt\int d\theta\Hquad \Phi^i\delta\Phi^i, 
$ is not manifest as in the pure bosonic case we discussed in the previous chapter. It is more suggestive, we first take the Berezin integral and then find a proper boundary condition that leaves $1D$  particle invariant under $\mathcal{N}=1$ SUSY transformations.
To see the calculations explicitly, let us take the abelian gauge group. By using the group manifold structure, one can always generalize to non-abelian results, starting by exploiting the flatness of $\mathfrak{F}$.
\begin{equation}
\mathcal{A}_t=\mathcal{G}^{-1}\partial_t\mathcal{G}=\partial_t\alpha+\theta\HHquad\partial_t \beta.
\end{equation}
Hence, the action becomes:
\begin{flalign}
    S^{\text{on-shell}}_{\text{Boundary-SBF}}=&\int_{\text{x=0}} dt \int d\theta\Hquad\Phi_i(t,0,\theta)\delta\left(\partial_t\alpha+\theta\HHquad\partial_t \beta)\right),\notag\\
    =&\int_{\text{x=0}} dt \int d\theta\Hquad\left(\psi\HHquad\delta\HHquad\partial_t\alpha+\psi\HHquad\theta\HHquad\delta\partial_t\beta +\theta\HHquad\phi\HHquad\delta\HHquad\partial_t\alpha+\theta\HHquad\phi\HHquad\theta\HHquad\delta\partial_t\beta\right)\label{variationofonshellboundaryBFaction},\\
    =&\int_{\text{x=0}} dt\left(-\psi\HHquad\delta\HHquad\partial_t\beta+\phi\HHquad\delta\HHquad\partial_t\alpha\right).\notag
\end{flalign}
In \eqref{variationofonshellboundaryBFaction}, we expand the $t$-component of the superconnection in terms of even and odd Grassmanns. In the second equality, we use \eqref{superfieldsevenoddstructure} and distribute the multiplication inside. Since both $\psi$ and $\theta$ are odd, their permutation will bring a negative sign, which is the reason for having that sign in front of the first term on the right-hand side of the last equality. By doing the same permutation on the last term in the second equality, we will have $\theta^2=0$ constraint on the unique Grassmann parameter. The form of the last equality is achieved by taking the Berezin integral.

\subsection{Manifestation of $\mathcal{N}=1$ supersymmetric quantum mechanics}
The previous section was reserved to find the form of the on-shell boundary action so that we can add a correction term that preserves the action principle as well as $\mathcal{N}=1$ SUSY.
Let us imply conditions on even and odd Grasmann's that live on the boundary $\{\text{x}=0\}$. 
\begin{flalign}
    \phi=&\partial_t\alpha,\quad\quad\quad\quad\psi=-i\beta.
\end{flalign}
We will have
\begin{equation}
    \int_{\text{x}=0}dt\Hquad \left(i\beta\HHquad\delta\HHquad\partial_t \beta+\partial_t\alpha\HHquad\partial_t\HHquad\delta\HHquad\alpha\right).
\end{equation}
In $1D$, the parity operator does not exist, while the only gamma matrix $\gamma^0=1$ is trivial. If one takes the variation out, one should add the $\frac{1}{2}$ for double counting of both Majarona fermions and bosonic part. 
\begin{equation}
    \int dt\Hquad\delta\left(\frac{1}{2}(\partial_t\alpha)^2+\frac{i}{2}\beta\partial_t\beta\right).
    \label{findingthecancelationequation}
\end{equation}
As one can see in \eqref{findingthecancelationequation}, if we realize the boundary limit of the fields as $\lim_{x\to 0}\beta=\chi$ and $\lim_{x\to0}\partial_t\phi=\dot{q}$.
\begin{equation}
    S^{\text{boundary}}_{\text{SBF}}=\int dt\Hquad\left( \frac{\dot{q}^2}{2}+\frac{i}{2}\chi\dot{\chi}\right),
\end{equation}
which is the $1D$ particle on the group quantum mechanics with a flat group manifold.\footnote{ For a generic gauge group $\mathbf{G}$ it would be
\begin{equation}
    S^{\text{boundary}}_{\text{SBF}}=\int dt\Hquad\frac{1}{2}\left( g_{ij}\dot{q}^i\dot{q}^j+ig_{ij}\chi^i\frac{D}{dt}\chi^j\right)
\end{equation}
where $\frac{D}{dt}$ is the covariant derivative of the group metric $g_{ij}$.} One needs to check whether this $1D$ particle on group quantum mechanical theory is invariant under SUSY transformations or not. For that purpose, the transformations below can be checked:
\begin{flalign}
    \delta q=&-i\epsilon\chi,\quad\quad\quad\quad\delta\dot{q}=-i\epsilon\dot{\chi},\notag\\
\delta\chi=&\epsilon\dot{q},\quad\Hquad\HHquad\HHquad\quad\quad\quad\quad\delta\dot{\chi}=\epsilon\ddot{q},
    \label{susytransformations}
\end{flalign}
where $\epsilon$ is the Grassmann odd supersymmetry generator. Now, taking the supersymmetric variation of the action, one finds
\begin{flalign}
    \delta_{\epsilon}S=&\int dt\Hquad \left(\dot{q}\HHquad \delta \dot{q}+\frac{i}{2}\delta\HHquad \chi\HHquad \dot{\chi}+\frac{i}{2}\chi\HHquad \delta\HHquad \dot{\chi}\right),\notag\\
    =&\int dt\Hquad \left(-i\dot{q}\HHquad \epsilon\HHquad \dot{\chi}+\frac{i}{2}\epsilon\HHquad\dot{q}\HHquad\dot{\chi}  +\frac{i}{2}\chi\HHquad\epsilon\HHquad\ddot{q}\right),\label{supercharge}\\
    =&\int dt\Hquad \left( -\frac{i}{2}\epsilon\HHquad\dot{q}\HHquad\dot{\chi}-\frac{i}{2}\epsilon\HHquad\ddot{q}\chi    \right),\notag\\
    =&\int dt\Hquad \frac{d}{dt}\left(-\frac{i}{2}\epsilon\Hquad\dot{q}\Hquad\chi\right).\notag
\end{flalign}
From the first equality to the second, we substituted the supersymmetric variations \eqref{susytransformations} inside. In the third equality, we used the Grassmann algebra properties to simplify it further. We write in total derivative form in the last equality and see that the extra factor is canceled. As it shows, we have a $\mathcal{N}=1$ transformation invariant $1D$ particle on the group prescription of quantum mechanics. Even though this theory is SUSY invariant and lives at the boundary of the $BF$ theory, seeing it as a "quantum" mechanics is a bit problematic. Real fermions on $\mathbb{R}^n$ lives in $2^\frac{n}{2}$ dimensional Hilbert space. This means SUSY quantum mechanics in odd dimensions will give an unsensible Hilbert space with $\sqrt{2}$ dimensions, which is not meaningful. However, again, following \cite{Michelson:1999zf} we can identify this specific $\{q,\chi\}$ supermultiplet as $\mathcal{N}=1\text{B}$, which is plausible since BF theory is equivalent to JT gravity and supersymmetric multiplet with real fermions are more relatable for black hole moduli spaces.

\subsection{$\mathcal{N}=1\text{B}$ $Osp(1|2)$ SUSY superconformal quantum mechanics}

We will see that our supergauge group governs the target space of our manifold that our particles defined on can be chosen as $Osp(1|2)$ with block structure as \cite{Britto-Pacumio:1999dnb}:

\begin{equation}\left[ 
\begin{array}{c|c} 
  Sp(2)\cong SL(2,\mathbb{R}) & \text{Fermionic} \\ 
  \hline 
  \text{Fermionic} & \mathbb{I} 
\end{array} 
\right]
\end{equation}
In our case, we can define the supercharge as 
\begin{equation}
    Q=\chi^a\Pi_a,
\end{equation}
where $\Pi^a=P_a-\frac{i}{2}\omega_{abc}\chi^b\chi^c$ and $\omega_{abc}$ is spin connection. One can again see that, upon the flat limit of the target manifold, we will recover the supercharge given in \eqref{supercharge}. The Hamiltonian of these real $Q^\dagger=Q$ supercharges can be written in terms of anti-commutator.
\begin{equation}
    \{Q,Q\}=2H
\end{equation}
such that by defining a supercovariant derivative $D=\frac{\partial}{\partial\theta}-i\theta\frac{d}{dt}\implies D^2=-i\frac{d}{dt}$ and $Q=\frac{\partial}{\partial\theta}+i\theta\frac{d}{dt}\implies Q^2=i\frac{d}{dt}$ while $\{Q,D\}=0$ the superspace action can be recast as
\begin{equation}
    S=\int dtd\theta\left(\frac{1}{2}g_{ab}\HHquad Dq^a(t,\theta)\dot{q}^b(t,\theta) \right),
\end{equation}
where $q(t,\theta )=q(t)-i\theta\chi(t)$. Berezin integral can be evaluated to find
\begin{equation}
    S=\int dt\left(\frac{1}{2}g_{ab}\HHquad\dot{q}^a\dot{q}^b+\frac{i}{2}\chi^a\HHquad g_{ab}\frac{D}{dt}\chi^b\right),
\end{equation}
where $\frac{D}{dt}\chi^a=\dot{\chi}^a+\dot{q}^b\Gamma^a_{bc}\chi^c$ is the covariant time derivative. One can see that choosing Christoffel symbols $\Gamma^a_{bc}=0$ on our curved group manifold and $g_{ab}=\delta_{ab}$ to a trivial group metric reduces our case. Since our spinors are gamma matrices $\gamma$ of Clifford algebra, our Hilbert space is a spinor \cite{Britto-Pacumio:1999dnb} where our covariantized conjugate momentum acts as a covariant derivative on the Hilbert space. By using the same homothety arguments as before, the $\mathcal{N}=1\text{B}$ superconformal quantum mechanics particle on $Osp(1|2)$ group algebra can be constructed from
\begin{flalign}
    H=\frac{1}{2}\{Q,Q\},\quad\quad Q=\chi^a\Pi_a,\quad\quad D=\frac{1}{2}D^a\Pi_a,\quad\quad
    K=\frac{1}{2}D^aD_a,\quad\quad S=i[Q,K]=\chi^aD_a.\quad\quad
    \notag
\end{flalign}
These suggest that our boundary theory particle on supergroup description is $\mathcal{N}=1\text{B}$ superconformal quantum mechanics with $\{q,\chi\}$ real spinor supermulitplet.

\section{Wedge-space holography for $\mathcal{N}=0,1$ 2D BF theory}\label{wedgesection}
Up to now, we have given the "simplest holography" construction for $SL(2,\mathbb{R})$ and $Osp(1|2)$ 2D BF theory. In this section, we will set up a wedge space holography picture mimicking gravitational holography in the sense \cite{Ogawa:2022fhy} with purely topological 2D BF theory. Given the web of dualities diagram:

\begin{equation}
\begin{tikzcd}[row sep=large,column sep=huge]
    \text{3\HHquad D CS theory} \arrow[d,"\text{dim. reduction}"] \arrow[r,leftrightarrow, "\text{holography}"] \arrow[dr, dashrightarrow,"\text{WSH}"]& \text{WZW} \arrow[d,"\text{dim. reduction}"]\\
    \text{2\HHquad D BF} \arrow[r, leftrightarrow,"\text{holography}"] \arrow[dr, dashrightarrow,"\text{WSH}"] & \text{1\HHquad D particle on group} \arrow[d, "\text{dim. reduction}"] \\
    & \text{0\HHquad D quantum mechanics}
    \label{diagram}
\end{tikzcd}
\end{equation}
 We will include the dashed lines with wedge space holography, i.e, a codimension 2 holography for wedges. We claim to have diagonal arrows that will correspond to physical theories. This will lead us to use a manifold with wedge space topology with a boundary and show that one can also find (anti)-chiral supermultiplets prescriptions of the Landau-Ginzburg type on the union space of the two wedges. To have a more complete picture, we will also present wedge space holography for $\mathcal{N}=0$ CS theory to 1d particle on group quantum mechanics in the appendix.
 \subsection{Wedge space for bulk 2d BF theory}
Now, let us have a BF theory defined on a wedge space $\text{M}=\{(t,\phi): -\phi^*\leq\phi\leq\phi^*\HHquad,t\in[0,\infty)\}$, where t runs like a radial coordinate. On this manifold, the boundary manifold corresponding to the end points of $\phi$ will be called $Q_1$ and $Q_2$, respectively. After finding the boundary theories live in those spaces, we will take the union space $Q_1\cup Q_2$ and check out the total theory. The boundary of the total theory will give us a codimension 2 relation with the original bulk theory. Unlike original construction with classical gravity on the bulk, we will have purely topological quantum theories that will bring out quantum to quantum codimension 2 holography structure. Let us define BF theory on this manifold with its original notation as Lagrange multiplier scalar $B$ field to avoid the confusions with angular coordinate $\phi$
\begin{equation}
    S_{\text{BF}}=\int^\infty_0 dt \int_{-\phi^*}^{\phi^*}d\phi\left(B\HHquad\partial_t A_\phi-B\HHquad\partial_\phi A_t\right),
\end{equation}
by following the same calculations in the previous sections
\begin{equation}
     \delta S_{\text{BF}}=\int^\infty_0 dt \int_{-\phi^*}^{\phi^*}d\phi\left(\partial_t(B\HHquad\delta A_\phi)-\partial_\phi(B\HHquad \delta A_t) -\partial_t B\HHquad \delta A_\phi+\partial_\phi B\HHquad \delta A_t  \right).
\end{equation}
 In our previous approach, we could safely claim the total t-derivatives vanish on the boundary, and the only non-trivial contribution is on \{\text{x}=0\} boundary. However, we now have two half-planes glued to give a wedge boundary, and they are parametrized as $\phi=[-\phi^*,\phi^*]$, $t\in[0,\infty)$. Considering this new boundary, we can proceed with our approach. First, let us assume that our fields are regular at $\text{t}=\infty$. To have a well-defined action principle, we should either impose well-defined boundary conditions or add external boundary terms that will act the same as our boundary theories on a wedge. If we have a flat connection, the equation of motion is satisfied on-shell, and the variation of the action will be in the form:
 \begin{flalign}
    \delta S_{\text{BF}}=&\Hquad \text{e.o.m}+\int_{-\phi^*}^{\phi^*}d\phi\int^\infty_0 dt \Hquad\partial_t(B\HHquad\delta A_\phi)-\int^\infty_0 dt \int_{-\phi^*}^{\phi^*}d\phi\Hquad\partial_\phi(B\HHquad \delta A_t)\notag,\\
    =&\Hquad \text{e.o.m}+\int dt \Hquad \left(B(t,\phi^*)\HHquad\delta A_t(t,\phi^*)-B(t,-\phi^*)\HHquad\delta A_t(t,-\phi^*) \right)\\+&\int_{-\phi^*}^{\phi^*}d\phi\Hquad B(0,\phi)\HHquad\delta A_\phi(0,\phi).\notag
\end{flalign}
Let us first focus on the second boundary term. In terms of the wedge space holography perspective, this term can be prevented with a UV cut-off. Bf theory, as one of the simplest Poisson-Sigma models, can be interpreted as a one-dimensional particle on group and Schwarzian mechanics. In the presence of this term, the boundary action promotes to Schwarzian mechanics since this term behaves like a string defect. By following \cite{Grumiller:2017qao}, we will simply impose the boundary along angular variable zero mode not varying along $\phi$. Hence $\int_{-\phi^*}^{\phi^*}d\phi\HHquad B(0,\phi)\delta A_\phi(0,\phi)$ where $ \delta A_\phi=0$. Now, by following the same calculations in previous sections, if we let a boundary cancellation term of the Schwarzian type on the action
\begin{equation}
S^{\text{BF}}_{\text{boundary}}=\int_{Q_1\cup Q_2} dt\Hquad \delta \left(\frac{1}{2}B^2(t,\phi^*)+\frac{1}{2}B^2(t,-\phi^*)\right),
\end{equation}
such that upon variation, it cancels the boundary term similar to \eqref{susyanalogueofschwarzianboundarycorrection}. Now, on the boundary upon variation, we have
\begin{equation}
    \delta S^{\text{BF}}_{\text{boundary}}=\int_{Q_1\cup Q_2} dt\Hquad \delta\left(\frac{1}{2}(\partial_t\varphi_1)^2+\frac{1}{2}(\partial_t\varphi_2)^2\right),
\end{equation}
where $\partial_t\varphi_{1,2}$ represents solutions of flat connections at $\phi=\phi^*$,$\phi=-\phi^*$ on $Q_1$ and $Q_2$ spaces where their union space runs with the same time t. Upon this construction, these scalar fields can be gathered under $\varphi=\varphi_1+i\varphi_2$ such that we have
\begin{equation}
 S^{\text{BF}}_{\text{boundary}}=\int_{Q_1\cup Q_2} dt\Hquad \delta\left(\partial_t \varphi^\dagger\partial^t\varphi\right),
\end{equation}
We have two different "particle on group" prescriptions on both spaces $Q_1$ and $Q_2$ with fields $\varphi_1$ and $\varphi_2$. Wedge space union $Q_1\cup Q_2$ allows us to write these fields under one complex scalar field such that we have a complex  1d particle on group with redefinitions $\partial_t\varphi\rightarrow \dot{z} $ and its complex conjugate $\partial_t\varphi^\dagger\rightarrow \dot{\bar{z}}$. Hence, our action becomes
\begin{equation}
    S_{\text{1d P.o.G}}= \int dt\Hquad \dot{\bar z}\dot{z}.
\end{equation}
By taking the holographic coordinate $t$ to zero to find the boundary of wedges, we will have 0-dimensional quantum mechanics with a complex charge.
\begin{flalign}
    \int^{\infty}_0 dt \Hquad\dot{\bar z}\dot{z}=\bar z(\infty)\frac{dz}{dt}(\infty)-\bar z(0)\frac{dz}{dt}(0),
\end{flalign}
where at infinity we assumed a vanishing charge.
\begin{equation}
    Q_{\text{0d charge}}= -\bar z(0)\frac{dz}{dt}(0),
\end{equation}
changing the topology of our holographic construction to the corresponding 1d particle on group prescription.
\subsection{$\mathcal{N}=1$ supersymmetric $BF$ theory on wedge}
In the previous case 2d $\mathcal{N}=1$ $ BF$ theory, we managed to find $\mathcal{N}=1B$ supermultiplet conformal quantum mechanics for our particle on group theory. In here, we will define our supersymmetric theory on wedge manifold $M$ and impose the same vanishing zero mode variation of superconnection $\delta\mathcal{A_\phi}=0$ on the angular boundary integral. Hence, starting from \eqref{SUSYBFTheoryaction1}  our new boundary will read as
\begin{equation}
    S_{\text{N}=1 \text{BF}}=\int dt\int d\theta\Hquad \left(\Phi(t,\phi^*)\delta\mathcal{A}_t(t,\phi^*)-\Phi(t,-\phi^*)\delta\mathcal{A}_t(t,-\phi^*)\right).
\end{equation}
After substituting $\Phi=\psi+\theta\phi$, $\mathcal{A}_t=\partial_t\alpha+\theta\partial_t\beta$ evaluated at respective points and taking the Berezin integral, we will have
\begin{flalign}
    S_{\text{N}=1 \text{BF}}=&\int dt\Hquad \left( -\psi(t,\phi^*)\partial_t\HHquad\delta\beta(t,\phi^*)+\phi(t,\phi^*)\partial_t\HHquad\delta\alpha(t,\phi^*)\right.\notag\\
    +&\left.\psi(t,-\phi^*)\partial_t\HHquad\delta\beta(t,-\phi^*)-\phi(t,-\phi^*)\partial_t\HHquad\delta\alpha(t,-\phi^*)     \right),
\end{flalign}
The chiral $\mathcal{N}=1$ supermultiplet with supergauge group Osp(1|2) can be seen by choosing
\begin{flalign}
    \psi(\phi^*)=&-i\beta_1,\quad\quad\quad\psi(-\phi^*)=i\beta_2,\\
\phi(\phi^*)=&\partial_t\alpha_1,\quad\quad\quad \phi(-\phi^*)=-\partial_t\alpha_1,
\end{flalign}
where t dependence is implicit and $\beta_{1,2}$, $\alpha_{1,2}$ evaluated at respective angular values as before. When we take the variation out and impose on-shell boundary action, upon variation, it becomes
\begin{equation}
     S_{1d P.o.G}=\int dt\Hquad \left(  
      \frac{i}{2}(\beta_1\partial_t \beta_1+\beta_2\partial_t\beta_2)+\frac{(\partial_t\alpha_1)^2}{2}+\frac{(\partial_t\alpha_1)^2}{2} \right),
\end{equation}
If we redefine group elements in complex parametrization setting $\xi=\beta_1+i\beta_2$, $\bar \xi=\beta_1-i\beta_2$ $\varphi=\alpha_1+i\alpha_2$
We achieve a particle on the group sigma model of a Landau-Ginzburg type with a complex boson and complex fermion chiral supermultiplet. If we let $\partial_t\varphi^\dagger\rightarrow \dot{\bar z},\HHquad\partial_t \varphi\rightarrow \dot z$ and $\xi\rightarrow \chi, \HHquad 
\bar \xi\rightarrow \bar \chi$
our boundary theory on $Q_1\cup Q_2$ becomes
\begin{equation}
    S=\int_{Q_1\cup Q_2}dt\HHquad \left(i\bar\chi\dot\chi+\dot{\bar z}\dot z  \right).
\end{equation}
This chiral supermultiplet is invariant under the following SUSY transformations \cite{Clark:2001zv}:
\begin{flalign}
\delta^Q\chi=&0,\quad\quad\delta^Q\bar\chi=-\dot{\bar z},\quad\quad\delta^{\bar Q}\chi=\dot z,\quad\quad\delta^{\bar Q}\bar\chi=0,\notag\\ \delta_Q z=&i\chi ,\quad\quad\delta^Q\bar z=0 ,\quad\quad\delta^{\bar Q} z=0,\quad\quad\delta^{\bar Q} \bar z=-i\bar \chi,
\end{flalign}
with the Lagrangian changing under SUSY as
\begin{flalign}
    \delta^Q\mathcal{L}&=\frac{d}{dt}\left(i\dot{\bar z}\chi\right),\quad\quad\quad\quad \delta^{\bar Q}\mathcal{L}=\frac{d}{dt}\left(-i\dot{ z}\bar\chi\right),
\end{flalign}
Moreover, one can see that this theory enjoys a global $U_j(1)$ invariance
\begin{flalign}
 \delta^{U(1)}z=iz,\quad\quad\delta^{U(1)}\bar z=-i\bar z,\quad \quad \delta^{U(1)}\chi=i\chi,\delta^{U(1)}\bar\chi=-i\bar\chi.
\end{flalign}
Hence, with wedge space holography, we change the topology of our manifold and achieve a chiral $\mathcal{N}=1$ chiral supermultiplet with $Osp(1|2)\otimes U_j(1)$ invariance. From now on, we can reduce the boundary of wedge space $Q_1\cup Q_2$ to a point, which will give:
\begin{equation}
    S^{1d P.o.G}=\int dt\left(\frac{d}{dt}(\bar z \dot z)+\frac{d}{dt}(\bar \chi\chi)-\bar z\ddot{z}-\dot{\bar \chi}\chi   \right),
\end{equation}
when equations of motion are satisfied we have $Q^{\text{0d QM}}=-\bar z(0) \dot z(0)-\bar\chi(0)\chi(0)$
as a single charge on a point.

\section{Conclusions }

In this work, we showed the "simplest holography" picture that is suggested by \cite{Janik:2018dll} can be extended to realize $SL(2,\mathbb{R})$ symmetry, the Schwarzian-type boundary condition also leads to a conformal quantum mechanics model on the particle on group description. Furthermore, we extend the general picture to "simplest supersymmetric holography" on $\mathcal{N}=1$, $2D$ $BF$ theory through dimensional reduction and showed that it is holographically equivalent to $1D$, $N=1B$ supersymmetric quantum mechanics where the deeper relations can be made by choosing our supergauge group as $Osp(1|2)$ with closed homothety that manifests a duality between $\mathcal{N}=1$ $2D$ $BF$/ $\mathcal{N}=1\text{B}$ \text{particle on Osp(1|2)} group superconformal quantum mechanics. Also, as a future outlook, we would like to explore other groups that admit these multiplets with references given \cite{Britto-Pacumio:1999dnb} and construct other quantum-to-quantum holographies from possible super-Schwarzian quantum mechanics \cite{Kozyrev:2021agn,Stanford:2017thb} that might manifest.  Moreover, our work extends the dualities diagram in \cite{Ebert:2022ehb} to $\mathcal{N}$=1 $BF$ $\leftrightarrow$ $\mathcal{N}$=1B superconformal quantum mechanics. This duality might lead to finding a relation between super-JT gravities \cite{Fan:2021wsb}, super-Chern-Simons theories and super-WZW theories to find a commuting dualities diagram with supersymmetry inclusion. The relation between Type B multiplets and black holes might be seen when we choose the gauge group as $\mathfrak{osp(1|2)}$. A wedge space holography with action formalism is constructed with the analogy of the ideas in \cite{Ogawa:2022fhy}. This brought out a complex particle on group description while in its supersymmetric extension, we got a Landau-Ginzburg type supersymmetric multiple and the supergauge group is extended with a global $U_j(1)$ invariance. By realizing wedge-space holography, we include diagonal reductions on the web of dualities diagram \eqref{diagram} and we believe that the relation between flat-space holography as codimension-$2$ holography, a quantum particle on a group prescription on the celestial sphere, might be constructed and insightful. Moreover, adding interactions to the quantum mechanical theories reduced from the holographies of this type is problematic. As a future outlook, we intend to see the wedge theory construction as a DCFT which might give interactive theories of the trace class of given particle on group manifold theories.

\section{Appendices}
\begin{appendix}
\subsection{$3D$ $CS$ theory on wedge}
Let us have a Chern-Simons theory on wedge space $\text{M}=\{(t,r,\phi): -\phi^*\leq\phi\leq \phi^*, r\in[0,\infty),t\in(-\infty,\infty)   \}$ where we intend to invoke wedge-space holography machinery. Boundary of $CS$ theory is well-known to be a BF theory, through wedge space holography construction we will show that its codimension $2$ holography will be a complex $1D$ particle on group quantum mechanics.
\begin{equation}
    S_{\text{CS}}=\int_\text{M} A\wedge dA=\int d^3x\Hquad \epsilon^{\mu\nu\rho}A_\mu\partial_\nu A_\rho.
\end{equation}
Now, let us rewrite this action in a more suggestive form for our purposes:
\begin{flalign}
    S_{\text{CS}}=&\frac{1}{2}\int_\text{M}d^3x\left( 
 A_t\left(\partial_rA_\phi-\partial_\phi A_r\right)+ A_r\left(\partial_\phi A_t-\partial_t A_\phi\right) +  A_\phi\left(\partial_t A_r-\partial_r A_t\right)\right)\notag,\\
 =&\int_\text{M}d^3x\left( A_r F_{\phi t}+A_t\partial_r A_\phi  \right)+\frac{1}{2}\int_\text{M}d^3x\left(\partial_t(A_\phi A_r)-\partial_r(A_\phi A_t)-\partial_\phi(A_tA_r)\right).
\end{flalign}
This dimensional reduction works for a general setting by choosing one holographic coordinate and a vanishing derivative \cite{Fan:2021bwt}. In our paper, we had a wedge space, for $2D$ $BF$ theory and our boundary theory on the face of the wedge space was $1D$ particle on group theory. In this case, our codimension $2$ holography will correspond to what we found for the face boundary of the wedge $BF$ theory and hence will diagonally close the diagram given in \eqref{diagram}. We will set $\partial_r=0$ while setting $A_r=B$ which mimics Schwarzian type boundary term that will reveal the $2D$ $BF$ theory on the face of the wedge space $3D$ $CS$ holography. We can do this construction safely since both the $CS$ theory and $BF$ theory have flat curvature. We will choose our holographic coordinate as $\phi\rightarrow\pm\phi^*$ and safely assume the total time derivative on the boundary will give regular fields after integration. Hence our $CS$ theory becomes:
\begin{flalign}
    S^{\text{reduction}}_{\text{CS}}=&\int d^2x\HHquad B F_{\phi t}- \frac{1}{2}\int dt\HHquad\left(B(\phi^*,t)A_t(\phi^*,t)-B(-\phi^*,t)A_t(-\phi^*,t)\right),\\
    =&\int d^2x\HHquad B F_{\phi t}- \frac{1}{2}\int dt\HHquad\left(B(\phi^*,t)^2+B(-\phi^*,t)^2\right),
\end{flalign}
where we choose $A_t(t,\pm\phi^*)=\pm B(t,\pm\phi^*)$ as before. When the equation of motion is satisfied this action principle will be valid hence $F_{\phi t}=0\implies A_t(t,\phi^*)=-\partial_t \phi_1(t,\phi^*)\text{  and  } A_t(t,-\phi^*)=-\partial_t \phi_1(t,-\phi^*) $  such that
\begin{equation}
    =\int d^2x\HHquad B F_{\phi t}-\int dt\HHquad\left(\partial_t\varphi^\dagger\partial^t\varphi\right),
    =\int d^2x\HHquad B F_{\phi t}-\int dt\HHquad\dot{\bar z}\dot z
\end{equation}
as in the previous calculation, we gathered two scalar fields that run with the same time $t $together at the union wedge space and produced a complex scalar field particle on the group. Furthermore, we identified them as quantum mechanics with complex Lagrangian
Hence, from the $3D$ $CS$ theory point of view, we have a wedge space holography with $2D$ $BF$ face boundary and complex scalar particle on group at the co-dimension $2$ holography limit.

\end{appendix}

\section{Acknowledgments}
The author expresses sincere gratitude to his dear colleague and best friend, Arda Hasar, as well as to Prof. Dr. Bayram Tekin, and Prof. Dr. Altug Ozpineci, Assoc. Prof. Soner Albayrak and Prof. Dr. Seckin Kurkcuoglu for their valuable discussions on the development of this work.

\end{document}